\documentclass[preprint]{revtex4}
\usepackage{bm}
\input{epsf}
\begin{document}
\title{A non--destructive analytic tool for nanostructured materials : Raman
and photoluminescence spectroscopy}
\author{A. Singha, P. Dhar and Anushree Roy }
\email{anushree@phy.iitkgp.ernet.in}
\affiliation{Department of Physics 
\\
Indian Institute of Technology, Kharagpur 721 302, WB, India}

\begin{abstract}
Modern materials science requires efficient processing and 
characterization techniques for low dimensional systems. Raman 
spectroscopy is an 
important non-destructive tool, which provides enormous information on 
these materials. This understanding is not only interesting in its own 
right from a physicist's point of view, but can also be of considerable 
importance in optoelectronics and device applications of these materials in nanotechnology.  The commercial Raman spectrometers are quite expensive. In this article, we have presented a relatively less expensive set-up with home-built collection optics attachment.  The details of the instrumentation have been described. Studies on four classes of nanostructures - Ge nanoparticles, porous silicon 
(nanowire), carbon nanotubes and 2D InGaAs quantum layers, demonstrate 
that this unit can be of use in teaching and research on nanomaterials.
\end{abstract}
\maketitle
\def\d{{\mathrm{d}}}
\section{Introduction}
In the last couple of decades, research on semiconductors has dealt with, 
quite literally, novel phenomena in lower dimensions - two, one and zero. 
In such systems, electrons are confined in planes, lines and mathematical 
points, respectively. The dimension of these materials in the direction of 
confinement is in nanometer scale, hence the name nanomaterials. With 
miniaturization, quantum effects become significant and need to be 
analyzed in explaining their unusual physical properties. Such systems 
are, therefore, inherently quantum in nature. The 2D systems are called 
quantum wells, the 1D systems - quantum wires and the 0D systems - 
quantum dots. Interestingly, because of quantum effects, the properties of 
such materials are significantly different from the properties of the same 
material at either single molecule or bulk scale. Modern technology 
exploits this fact and uses these materials in diverse applications. For 
example, bulk CdS is yellow in colour. By changing the size of the 
nanoparticles of CdS, one can tune its colour. These 'coloured' particles 
can then be used to tag biological molecules. Thus, nanoscience in 
conjunction with bioscience, can meet the challenge in scanning and 
identifying through 
thousands of genes and millions of proteins in search of new drugs and 
drug targets\cite{Han:2001}. Research and development sectors all over the 
world embraced the use of nanomaterials based on silicon (Si) and 
germanium (Ge) for 
new and efficient optoelectronic device fabrication \cite{Hirschman:1996}. 
A new allotrope of carbon, the carbon nanotube (also called buckytube) has 
become a vital component  in the fabrication of transistors 
\cite{Collins:2001} and flow sensors \cite{Ghosh:2003}. Apart from 
optoelectronics and bioscience, where nanomaterial applications are in 
abundant use, nanostructured materials are also important in information 
technology, integrated complex products such as hard disk drives that 
store information, silicon integrated circuit chips that process 
information in every Internet server and personal computer and in defence 
related applications 
\cite{Ouyang:2001,Ahmed:1996,Grethlein:book}. There are 
others too. For example, a nanocomposite of epoxy layered silicate, can be 
employed as the primer layer of corrosion prevention coating in aircrafts 
and other high performance vehicles \cite{Chen:2003}. Thus, these new 
state of art materials play a pivotal role in facilitating improved as 
well as economic performance of a broad spectrum of instruments and 
devices, thereby uplifting the quality of life. A great emphasis 
is therefore needed in order to understand the fundamental properties of 
these low dimensional systems. Many research programmes, both in basic 
science and industry, have been initiated around the world over the last 
few years in an intense effort to harness the potential of these materials 
for the generation of new  technology. 

Scientists need new and powerful analytic techniques to study these 
nanostructured materials, to discover the novel properties of these systems 
and to improve the materials for the future. Tools such as high-resolution 
transmission electron microscopy, 
scanning probe microscopy, x-ray diffractometry, photoluminescence 
spectroscopy and Raman and infra-red (IR) spectroscopy can be used to 
study different characteristics of these materials.  Raman and IR 
spectroscopy measure vibrational levels of the molecules. These two 
techniques can be used complementarily, because, due to the differences in 
spectroscopic selection rules each is sensitive to different 
components of a given centrosymmetric molecule. For noncentrosymmetric 
materials these two techniques probe the same vibrational levels of  
molecules. However, there are many advantages of Raman spectroscopy over 
IR measurements - i. no tool-specific sample preparation is required 
(Nujol or KBr matrices are not used in Raman spectroscopy). Incident 
radiation can directly interact with the sample, ii.  Raman does not 
suffer the material limitations of infrared spectroscopy since both glass 
and water exhibit minimal Raman spectral interference, iii. Raman bands 
are typically narrower than those observed in mid-IR spectra and can be 
used more readily for quantitative analysis, a fact to be discussed later 
in this article. Finally, Raman spectroscopy has a particular importance 
in the area of nanomaterial synthesis and their application in technology. 
This technique can therefore be uniquely placed to support the growing 
demand of understanding the properties of the novel materials of today.

In section II of this article, we have first discussed Raman scattering and 
Photoluminescence phenomena in brief along with how these two techniques 
can provide new and detailed information on the properties and quality of 
nanostructured materials. Section III describes the details of the 
spectrometer, which we have assembled, to carry out the above studies. 
Section IV deals with the calibration of this spectrometer and the study 
of standard samples. Using the above instrument, Raman and 
photoluminescence spectroscopic studies on different nanostructured 
materials are discussed in section V. Finally, in section VI  we summarize 
our results and make a few concluding remarks.


\section{Raman and photoluminescence spectroscopy : basics}

\subsection{Raman spectroscopy} 

Inelastic scattering of light was first predicted in 1923 by A. Smeckal 
\cite{Smekal:1923}. However, it was not until 1928 that Sir C.V. Raman 
carried out the first set of experiments, which confirmed the prediction 
and led to the Nobel award in 1930 \cite{Raman:1928}. Unlike Rayleigh 
scattering \cite{Rayleigh:1899}, which is elastic scattering of light, 
Raman scattering is an
inelastic process caused by some quasi excitations of the medium. 
These excitations can be vibrational modes in a molecule, phonons in a
crystal, plasmons, single particle electronic excitation, magnons etc 
\cite{Hayes:book,Cardona:book}. 

\begin{figure}
\centerline{\epsfxsize=2.25in\epsffile{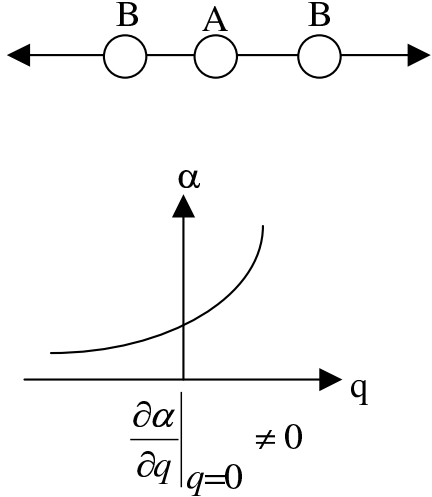}}
\caption{Dependence of $\partial \alpha/\partial q$ on 
the normal coordinate of vibration for a AB$_{2}$ type molecule}
\end{figure} 

Classically, Raman scattering can be explained by a time varying
polarizability of the molecule modulating an optical response. When a
molecule is subjected to the electric field ${\bf E} ={\bf E}_{0}cos 
{\omega}t$ of a beam of electromagnetic radiation, the dipole moment {\bf p}
of the molecule is given by
\begin{equation}
{\bf p}={\bf \mu}_{0}+{\tilde \alpha}{\bf E}
\end{equation}

\noindent
where ${\bf \mu}_{0}$ represents the permanent dipole moment, while
${\tilde \alpha}{\bf E}$
is the induced dipole moment. In general, the polarizability is 
represented by a rank two tensor, $\alpha_{ij}$. The structure of 
$\alpha_{ij}$ depends on the symmetry of the molecule. For small
vibrations the normal coordinates $q_{n}(t)$ of the vibrating
molecule can be approximated as $q_{n}(t)=q_{n0}cos(\omega_{n}t)$, where
$q_{n0}$ is the amplitude and $\omega_{n}$, the vibrational frequency
of the $n$th normal mode. The total dipole moment is then given by

\begin{eqnarray}
{\bf p}={\bf {\mu}_{0}}+\alpha_{ij}(0)E_{0}cos(\omega{t}) 
+\sum_{n=1}^{Q}\left(\frac{\partial{\bf
{\mu}}}{\partial{q_{n}}}\right)_{0}
q_{n0}cos(\omega_{n}t)\nonumber \\ +
\frac{1}{2}{\bf
E}_{0}\sum_{n=1}^{Q}\left(\frac{\partial{\alpha}_{ij}}{\partial{q}_{n}}\right)_{0}
q_{n0}
[cos(\omega+\omega_{n})t+cos(\omega-\omega_{n})t]
\end{eqnarray} 

\noindent
where the Taylor series expansions in $q_{n}$, for $\mu$ and $\tilde 
{\alpha}$ have been used and only first order terms retained.
The second term describes Rayleigh scattering, the third term the 
Infrared spectrum and the fourth term, the Raman scattering 
process. Thus, for a mode of vibration of a molecule to be 'Raman 
active', the necessary criterion is 
$\frac{\partial{\alpha}}{\partial{q}}{\vert}_{q=0}\neq{0}$ 
[Fig. 1].

\begin{figure}
\centerline{\epsfxsize=3.25in\epsffile{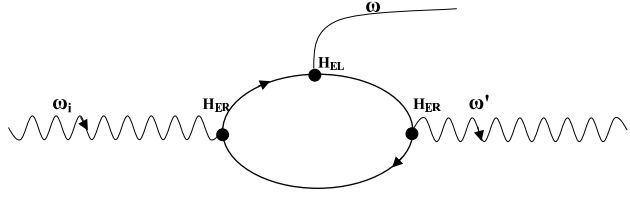}}
\caption{Schematic of Raman scattering process}
\end{figure}  

In microscopic theory, inelastic 
scattering of an incident photon by phonons, for example, in a crystal is 
defined by a third order interaction process between electron and 
radiation (denoted by the interaction Hamiltonian H$_{ER}$), electron and 
lattice (H$_{EL}$) \cite{Loudon:1964}.  Fig. 2
illustrates the Raman scattering process in terms of these elementary 
interactions. An optical photon (frequency $\omega_i$ and wavevector 
${\bf k_i}$) cannot 
interact with the lattice directly; rather it interacts with an electron 
via the electron radiation interaction. An electron-hole pair is created 
in this process. After creating (or annihilating) a phonon of frequency 
$\omega$ 
(and wavevector {\bf k}) via electron-lattice interaction, this electron 
recombines with the hole via electron-radiation interaction. The 
interaction vertices (denoted by the black dots in Fig. 2) makes the 
overall process of third order (i.e. second order in the 
electron-radiation interaction and first order in the electron lattice 
interaction). If the 
scattered phonon frequency is $\omega_{s}$ (and wavevector ${\bf k_s}$), 
energy and 
wavevector conservation lead to
\begin{equation}
\omega_{i}=\omega_{s}\pm\omega \hspace{0.2in}{\bf 
k_{i}}={\bf k_{s}}+{\bf k}
\end{equation}
 
\noindent
If $\omega_{s}$ is smaller (larger) than incident photon frequency, 
$\omega_{i}$, it is 
called Stokes (Anti-Stokes) scattering. Change in the incident photon 
frequency (often expressed in wavenumber, cm$^{-1}$), $\omega_{i}-\omega$, 
is called the Raman shift.

The Raman scattering cross-section can be shown to be  proportional to (i) 
$\omega_{s}^{4}$ and 
(ii) the second order term of the polarizibility tensor 
$(\alpha_{ij,n} = 
{\left[\frac{\partial\alpha_{ij}}{\partial q_{n}}\right]}_{q=0}$) 
associated with 
the electrons in the material \cite{Loudon:1964} : (i) implies that for 
smaller wavelength the spectral lines are more intense. It is also easy 
to note that because of (ii) Raman scattering is a very weak process and 
therefore not easy to detect without specific experimental arrangements.  

Unlike bulk systems, phonons are typically confined in 
nanomaterials. As we shall see below, for low dimensional systems, Raman 
spectroscopy can be used to investigate these confined and 
surface/interface  phonon modes. In addition, it can also be used to study 
structure, quality and thermodynamic properties of materials. We now 
briefly outline  these aspects.

{\em Confined phonon modes} : An acoustic phonon in a bulk crystal is 
usually described 
by the vibrational theory of a continuous elastic body (all the atoms 
in a crystal move 
together, as in a long wavelength acoustic vibration). On the other 
hand, in zero 
dimensional systems, these modes arise due to elastic vibrations of a 
homogeneous elastic body as a whole (that is why it is also called 
particle mode).  Under stress free boundary condition, it gives rise to 
two acoustic modes : spheroidal (vibration with dilatation) and torsional 
(vibration without dilatation). Spheroidal modes are Raman active and the 
frequencies of these modes are inversely proportional to the size of the 
particles \cite{Lamb:book,Duval:1986, Roy1:1996}. 
Because of very low frequency the acoustic modes for the bulk 
systems cannot be observed by Raman measurements, but for low 
dimensional systems they appear in the measurable frequency region 
(below ~100 cm$^{-1}$) of the Raman spectrum.
Furthermore, in two dimensional systems, we observe a new phenomenon - the 
folding of acoustic phonon modes. A 2D quantum well structure is formed by 
layered growth of two semiconductor species with different band gap 
energies. Acoustic phonons in such a structure
overlap in frequency and
can propagate normal to the interface, i.e. along the growth 
direction. The folding of acoustic 
phonon modes in a layered 
structure is understood using either the (i) linear chain model (LCM) 
\cite{Colvard:1985} or (ii) 
the effective continuum model (ECM) \cite{Rytov:1956}. We briefly describe 
(i) here. Imagine a linear chain of 
atoms representing the layered structure across the interface. For, 
example, in a Ga-As, Al-As layer the chain contains alternating Ga, As and 
Al atoms. The elastic coupling between Ga, As is obviously different from 
that for Al, As. The set of equations governing the displacements of each 
atom forms a coupled oscillator system. For individual layers one can work 
out the solutions. Across each layer of different constituents one needs 
to match the forces. The end result of such matching reveals the 
folding of the phonon modes at the new zone boundary, $\pi$/d,
where d=d$_1$+d$_2$
(thickness of the layers of the components). This is 
schematically shown in Fig. 3. In (ii) one assumes an elastic continuum 
with different sound velocity and density on either side of the boundary 
separating the layers. The solution of the wave equation and its 
derivative are matched at the boundary. The folding of acoustic phonons 
can also be explained in this 'continuum' framework, where the qualitative 
result remains the same as in LCM. These typical features of 
the confined acoustic phonons in low dimensional systems can be observed 
and quantified successfully through Raman measurements. The interested 
reader can look up \cite{Sood:1989} for a lucid discussion 
on these aspects.
\begin{figure}
\centerline{\epsfxsize=2.25in\epsffile{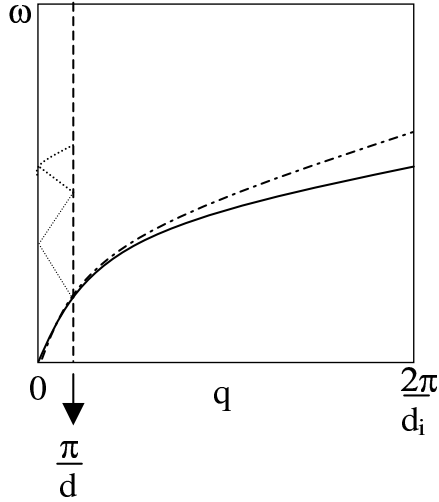}}
\caption{Schematic illustration of the folding of acoustic phonon (dotted
line) within mini zone boundary, $\pi$/d (dashed line). The bold line and
the dashed-dotted lines
are the acoustic phonon dispersion curves ($\omega$- q, i=1,2) for the
two  components of the layered structure}.
\end{figure}

In contrast to the acoustic phonons, the optical phonon frequencies  of 
the two constituent materials of the layered structure may overlap in some 
region or 
may not overlap. In the former case, the optical phonon modes can be 
described 
by folding into the  new mini zone boundary, $\pi$/d, similar to acoustic 
phonon  modes. In the latter case, phonons cannot propagate in the growth 
direction, hence they are confined in each layer. Confinement of optical 
phonon modes in low dimensional systems will be discussed later in 
section V of this article. We note here, 
that the propagation of acoustic phonons is sensitive to the periodicity 
of the well-structure, while optical phonons are sensitive to interface 
properties of the each individual quantum well. 

{\em Interface and surface phonons} : As mentioned before, Raman 
spectroscopy can be 
used to study interface and surface phonons in nanomaterials 
\cite{Fuchs:1968,Roy2:1996,Ruf:book}. 
The long wavelength optical modes, which correspond to  coupled 
excitations of  phonons and photons, are also known as polaritons. Their 
frequencies are complex, the imaginary parts arise from both anharmonicity 
and radiative damping. Detailed analysis shows \cite{Fuchs:1968}
that the polaritons of longitudinal electric type are bulk modes, 
whereas the polaritons of
transverse magnetic type are surface modes. From 
the coupling of the electric field of the 
photon to the dielectric polarization of the transverse optical phonon, 
the dielectric function is obtained as
\begin {equation}
\epsilon(\omega)=\epsilon(\infty)\left(\frac{\omega_{L}^{2}-\omega^{2}}
{\omega_{T}^{2}-\omega^{2}}\right)
\end{equation}
$\omega_{L}$ and $\omega_{T}$ are longitudinal optical (LO) and transvere 
optical (TO) phonon frequencies, respectively. $\epsilon(\omega)$ 
is negative  when $\omega_{LO} < \omega < \omega_{TO}$.
For a plane wave propagating in 
the x-direction in a bulk crystal, the temporal and spatial variation of 
the wave is described by the factor $\exp{[i(kx-\omega t)]}$. In the 
frequency range 
for which $\epsilon(\omega)< 0$, $k$ is imaginary 
($k=\sqrt{\epsilon(\omega})$) and the wave decays 
exponentially. 
Therefore, in this frequency range, an electromagnetic wave cannot 
propagate much in the bulk crystal. These surface modes in the bulk 
sample are so weak that 
they cannot be detected by Raman measurements. In contrast, because of 
high surface to volume ratio, these modes (which involve interface 
modes as well) appear in low dimensional 
systems and are detected easily by Raman scattering.  

{\em Structure and quality of the material} : Raman measurements also  
provide 
detailed information on 
the structure and quality of low dimensional materials. The reason behind  
is that lattice vibration is very sensitive to nearest neighbour 
interaction and therefore can efficiently probe crystal structure and 
quality in extremely small scale (of the order of lattice spacing). For 
example, investigations on longitudinal optical (LO) phonons in 
semiconductors via Raman spectroscopy provide significant information 
about the strain, defects and disorder in these materials 
\cite{Jiang:1999,Roy:1994}. Measurement of the Raman spectrum in 
these systems tells us about the microstrain in semiconductors under 
stress. This inherent stress at the interface causes the polarization 
dependent splitting 
and/or shift of the Raman line, which vary linearly with the stress. Raman 
shift due to the hydrostatic component of the strain yields a value for 
the Gr\"{u}neisen parameter (average value of change in vibrational 
frequency 
of the atoms in a crystal per unit dilation) for the new material 
\cite{Cerderia:1972,Liarokapis:1999}. 

{\em Phase transitions}: An additional thermodynamic 
aspect is phase transitions, which can  play an important role in device 
fabrication. For instance, the current injected laser diodes based on GaN 
require suitable thermodynamic properties for their stability and 
resistance to degradation under conditions of high electric currents 
and 
intense light illumination \cite{Karpinski:1984}. Phonons are primary 
excitations, which influence the thermodynamic behaviour. The pressure 
dependence of the optical mode energies can be related to homogeneous 
shear strain caused by a phase transition \cite{Decremps:2002}. This shear 
strain in the crystal can be measured from the change in frequency of the 
Raman shift. The phase transitions in a material under different 
conditions 
can be identified from the discontinuities in the Raman peak positions, as 
well as from the changes in the number of observed Raman lines 
\cite{Chandrabhas:1995}.

In summary, the crucial knowledge 
of strain, disorder, phase in a material can be obtained with a fair 
amount of precision through specific features of the Raman shift.

\subsection{Photoluminescence} 

\begin{figure}
\centerline{\epsfxsize=3.5in\epsffile{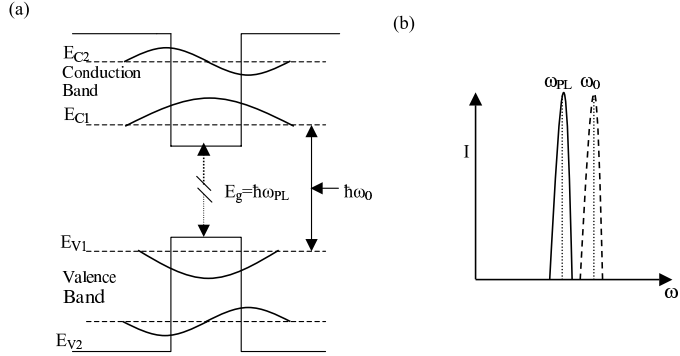}}
\caption{(a) The electronic transition between the quantized levels 
E$_{c1}$ to 
E$_{v1}$ in a semiconductor low dimensional system in comparison with 
PL energy of the bulk system (b)
corresponding photoluminescence intensity  as a function of frequency for 
low dimensional system (dotted line) and for the bulk system of same 
material (solid line)}
\end{figure}

The above discussion tells us that Raman scattering essentially probes the 
electron-lattice interaction. On the other hand, photoluminescence is 
directly related to electronic structure and transitions.
Differences in the electronic behaviour between bulk and low 
dimensional semiconductors arise due to the difference in the electronic 
density 
of states. For example, density of states of electrons, N(E), of a bulk 
semiconductor has a continuous dependence N(E)$\sim$E$^{1/2}$, where-as 
in 2D 
semiconductors it has a step-like dependence with energy. In 1D 
semiconductor structures it diverges with energy as E$^{-1/2}$  and for 0D 
semiconductor it is a delta function of energy. 
In semiconductors, electrons in the 
valence band absorb photons and get lifted into the conduction band 
everywhere, leaving holes in the valence bands. In a bulk sample, this 
occurs provided $\hbar \omega$$\geq$E$_g$, where
E$_g$ is the band gap of the semiconductor and $\omega$ is the
frequency of incident photon. Some of these excited 
electrons lose their energy by nonradiative transition.  It is then 
possible for an electron to fall from conduction band into the holes in the 
valence band by a radiative transition process. Usually the lowest 
levels are seen in PL spectrum and the emitted PL 
frequency, $\omega_{PL} < \omega$. 
Thus, PL spectrum of a bulk semiconductor, in general, corresponds to 
its band gap. 
In low dimensional systems, because of the quantization of energy levels 
in the conduction and valence bands (increase in bandgap) the frequency of 
PL emission, $\omega_{0}$, corresponds to the transition of excited 
electrons from the 
lowest state in the conduction band (E$_{C1}$) to the lowest state in 
the valence band (E$_{V1}$) [Everthing is upside down in the valence 
band, as shown in Fig. 4. Since $\omega_{PL} < \omega_{0}$
PL peak in these systems shifts from that in corresponding bulk materials. The 
schematic of this phenomenon is illustrated in Fig. 4.
The process can 
be direct or indirect depending on the gap energy. The characteristic 
shift of the band gap with decrease in size of the clusters can be studied 
from the blue-shift in PL peak position \cite{Moriarty:2001}.  This 
technique has been shown to be useful in the study of quantum confinement 
of electrons in low dimensional systems.  

Due to change in band gap some 
of these nanostructured materials can have tunable light emission, where 
the tuning depends on the size of the nanostructures. These have many 
potential applications in optoelectronic devices \cite{Hirschman:1996}. PL 
spectroscopy also yields information on the surface state density through 
intensity variations \cite{Kim:1967} and width of the spectrum. Surface 
states are caused by the interruption of the periodical arrangement of the 
atoms and by the deposition of impurities at the surface. Because of large 
surface to volume ratio this effect is more pronounced in nanoparticles. 
Thus, PL spectroscopy offers a tool for improving our understanding of the 
compound semiconductor surface. Moreover, for device fabrication using 
semiconductor low dimensional systems, such studies demonstrate the 
feasibility of improved technology. For example, the efficiency of solar 
cells is limited by non- radiative recombination occurring in the bulk via 
defect states or at the surfaces via surface states. PL spectroscopy also 
gives us information on band bending \cite{Lester:1986}, surface related 
transitions \cite{Kim:1987} and  near-surface bulk properties 
\cite{Bebb:book}.

\begin{widetext}
\begin{table}
\begin{tabular}{|c|c|}\hline
{\bf Raman spectroscopy} & {\bf Photoluminescence spectroscopy} \\ \hline 
\hline
Interface properties of  individual quantum well
& Origin of luminescence in quantum systems \\ \hline
Periodicity of the quantum well structure
& Electronic states of the materials\\  \hline
Size of the quantum dots from the confinement of phonons
& Electron hole exchange interaction \\ \hline
Local field effects in quantum dots embedded in a matrix
&Surface state density -provides information for \\
& compound semiconductor surface \\ \hline 
Defect/disordered states in the quantum structures &
Surface related electronic transition \\ \hline
Microstrain in the quantum systems & Electronic band bending
\\ \hline
Gr\"{u}neisen parameter of the low dimensional systems
& Near surface bulk properties.\\ \hline
Thermodynamic properties, like phase transition &
\\ in the materials & \\ \hline
Dopant concentration in & \\ semiconductor quantum dots &  \\ \hline
\end{tabular}
\caption{Brief listing of information on the nanomaterials obtained from
Raman and PL spectroscopy}
\end{table}
\end{widetext}

Thus, Raman and PL spectroscopy yield structural and 
dynamic information on materials at the molecular level. Unfortunately, 
modern commercial spectrometers are highly expensive and becoming an 
option for the budget in many research labs.  In this paper, we present 
the design of a relatively low cost Raman spectrometer, which can ideally 
be used to study today's state of art nanomaterials. Table I briefly 
summarizes how Raman and PL spectroscopy can be used to study 
nanomaterials.

\section{Instrumentation}
As we mentioned before, being a third order interaction process Raman 
scattering is an extremely weak phenomenon. Moreover, because of very low 
volume fraction of the materials in nanostructured systems, it is often 
quite difficult to get a Raman signal without damaging the sample by a 
high power laser. Thus, one needs a clever choice of the optics to collect 
scattered light from the sample.  The low cost spectrometer, described  
here is useful both for Raman and PL measurements. The instrument consists 
of an inexpensive single Monochromator with open electrode CCD detector 
and 100 mW air cooled Argon ion laser source. The collection optics is 
home-built. Schematic diagram of the spectrometer is shown in Fig. 5. The 
details of the instrument is discussed below.

\begin{figure}
\centerline{\epsfxsize=3.5in\epsffile{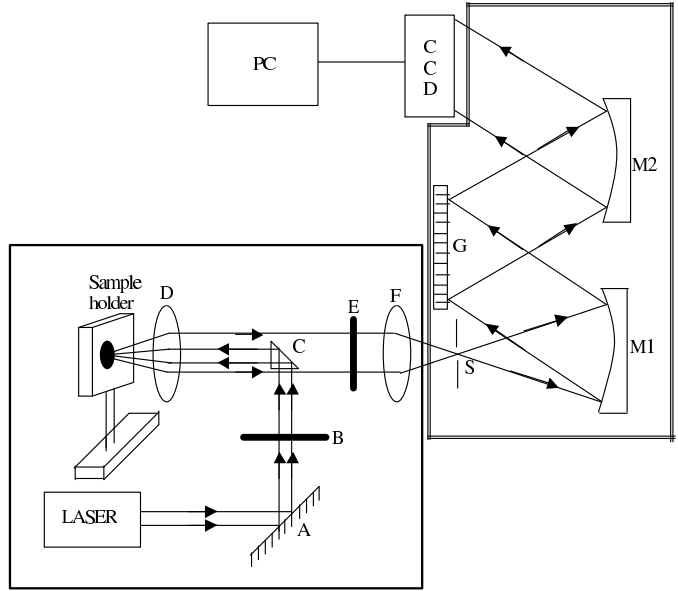}}
\caption{Schematic diagram of the Raman set-up.  The actual size of the 
prism (C) is much smaller.}
\end{figure}

{\em Excitation source} : In the pre-laser era the main drawback was the 
lack of 
a sufficiently intense radiation source to detect the weak signal from 
inelastic scattering.  Introduction of lasers has revolutionized this 
classical field of spectroscopy. To get detectable scattering intensity 
one usually uses either water cooled high power (Watt) lasers, which are 
not only expensive, but also difficult to maintain. One needs good 
infrastructure like stable power and water supply to run it efficiently. 
We have used a single line polarized (TEM$_{00}$ mode) 100 mW air 
cooled argon ion laser of wavelength 488 nm, 
Model no. 177-G12 from Spectra Physics, as an excitation source. Because 
of vacuum sealing and factory aligned cavity optics, an air-cooled laser 
does not require any adjustment or regular cleaning; this eliminates 
costly system downtime. We have used  a 5KV UPS  for stable power supply 
for the laser and we obviously do not need constant water circulation to 
run this 
laser.  This laser is economic compared to a high power laser, easy to 
maintain and it serves our purpose. 

{\em Collection Optics}: In the usual commercial set-ups, low power (mW) 
lasers are used along with a
MicroRaman system to record the weak Raman scattering signals. The 
collection optics in such a set-up consists of expensive
microscope objective along with a notch filter\cite{Voor:1994}. MicroRaman 
spectrometer also requires
crucial alignment of the Monochromator and detector. In contrast, 
in our set-up, except a notch filter, we have
used simple in-expensive optical components (mirror, lens and prism) for
collecting the Raman signal from the sample and directing it to the
Monochromator. The schematic diagram of the  collection optics is shown 
in the square marked (single line) area in Fig. 5.  It is easy to align 
all the components for collection of
the spectrum. For stability and alignment of the system we have kept the
optical components on a breadboard. Using a mirror (A), light from the
laser is steered to plasma filter (B) [Omega optical Inc.] and then to a
tiny right angle prism (C), where it bends through 90 degree.  This light
is focused on the sample by a Nikon camera lens (D) of focal length 55 mm.  
The scattered light is collected by the same lens, then it passes through
a holographic Super Notch filter (E) [Model no. 1495-488-SNP from Keiser
Optical system Inc.]. This notch filter is used to cut off the Rayleigh
line from the scattered light. After passing through the notch filter the
scattered light is then imaged on the entrance slit S of the Monochromator
using a collecting and focusing lens (F) of focal length 55 mm.

For high efficiency of the 
instrument the selection of the above optical components has been made 
very carefully based on detailed estimation of several parameters, in 
particular - focal length of the lens, size of the prism, 
reflectivity of the mirror. We have used high quality front face 
reflecting mirror (A) of reflectivity 98$\%$  at 488 nm for steering of 
the beam. We have chosen a very tiny prism (C) of flint glass 
and appropriate holder so that the scattered light from 
the sample (1 in $10^5$) is not hindered. The Abbe number of the prism is 
36.37. Inverse of Abbe number is called the dispersive power 
(does not play a vital role in our set-up) of the prism.
In our system, the scattered light is collected in 180 degree scattering 
geometry. This set-up does not require any vibration isolation table. A 
heavy wooden table is providing a reasonable stability to the set-up by 
eliminating spurious vibrations appreciably.

{\em Notch filter} : In the scattering process strong elastically 
scattered 
Rayleigh background often masks weak inelastically scattered Raman 
signals. To remove this Rayleigh background, usually high-cost double or 
triple Monochromator is used as dispersive unit. These Monochromators are 
expensive and a large number of optical components in the dispersive unit 
attenuate the scattered light quite significantly. To get the appreciable 
intensity of the inelastically scattered light it is essential to use high 
power lasers as an excitation source with double/triple Monochromators. 
The background due to Rayleigh scattering cannot be suppressed efficiently 
by a single Monochromator, especially at low wavenumbers. But, the 
combination of single stage Monochromator with  Rayleigh line rejection 
filter, i.e Notch filter, of large optical density at the laser frequency 
offers an efficient and low cost alternative. The super notch filter, 
which we have used, has an optical density  $>$6.0 at 488 nm, spectral 
band 
width of 350 wavenumber and spectral edgewidth of $\sim$ 150 wavenumber. 

{\em Monochromator} : We have used a single stage ISA TRIAX 550 
Czerny-Turner 
Monochromator as the dispersive unit. The schematic diagram of the optics 
of grating Monochromator is shown in the square marked (double lines) area 
in Fig. 5. The slit S is placed in the focal plane of a concave 
mirror M1.  Mirror M1 and M2 image the entrance slit on the detector. It 
uses 76mm $\times$ 76mm holographic grating (1200 rulings/mm) for 
dispersion of 
the spectral line. The density of the groove at longer wavelength provides 
the best resolution of the spectrometer as 0.025 nm. The large grating 
provides very good light collection with f/6.4 aperture. The width of the 
slit S can be varied from 2 $\mu$m to 2000 $\mu$m. This Monochromator 
has focal length 0.55 m. It can be used for both Raman and PL 
measurements. 

{\em Detection system} :   To measure very weak  signals Charge Coupled 
Device 
(CCD) is often used as a detector at the exit slit of the Monochromator of 
a Raman set-up. Multichannel CCD helps one to collect more scattered light 
signals for a wide spectral window in shorter time than what can be done 
by a Photomultiplier tube (PMT). Thus, using this device a very weak 
signal can be collected with help of a very long integration time. 
Depending on design and operating parameter there are three kinds of CCD 
detectors available - i. Front illuminated ii. Back thinned iii. Open 
electrode.  The optimum choice of a CCD detector depends upon i. 
wavelength range of interest ii. amount of anticipated signal iii. 
spectral coverage and resolution. We have used high performance 
thermoelectrically cooled, of 1024$\times$256 pixel, front illuminated 
open 
electrode (OEL) CCD chip. Though the quantum efficiency of this kind of 
CCD detector is only 53$\%$ in the visible and near infra-red (NIR) range 
(400 nm - 900 nm) and 30$\%$ in UV region compared to quantum efficiency 
of  
90$\%$ (only in visible range, starts to fall off above 800 nm and below 
450 
nm ) in back thinned CCD chips, we have chosen this particular kind of 
detector for the following reasons: A) unlike back-thinned and front 
illuminated CCD detectors  the dynamic range of OEL CCD is not a limiting 
factor as it has an average quantum efficiency of 40$\%$ with relatively 
large response over 200-900 nm. The spectrometer with this detector can be 
easily used to study PL  in different semiconductor materials, Raman 
scattering from biological samples (which usually needs detection of light 
in infra-red/NIR regions) using appropriate laser source. Moreover, though 
the quantum efficiency of OEL CCD is comparatively low, one can overcome 
this 
drawback by collecting weak signals with long integration time, B) when 
back thinned CCDs are used in NIR region, etalonic effect appears due to 
interference induced by reflection of some wavelengths at the boundaries 
of the silicon layers in the chips. This results in oscillations 
superimposed on collected CCD data. Etalonic effect can be minimized by 
reducing the size of the image focused onto the detector or by changing 
the central wavelength of the acquisitions. However, this effect cannot be 
totally suppressed. With OEL one can completely avoid such problem, C ) 
OEL CCD detectors are cheaper (nearly half the price) compared to 
back-thinned CCD detectors.

We believe that OEL CCD detectors are good enough for Raman and PL 
measurements, which we are aiming at.  We have measured very weak Raman 
signals from nanomaterials using this detector (discussed later).

{\em Interfacing, data acquisition and analysis}: Data acquisition is done 
by 
interfacing CCD and computer via IEEE card no. P/N 973027. For data 
acquisition and data manipulation we have used a Windows based programme 
'Spectramax'\cite{Spectrmax:os}. This commercial programme is also capable 
of fitting the Raman peaks by non-linear curve fitting programme, data 
editing, conversion of units, calculation of peak area and line width. 
Using the same software one can also collect raw data in ASCII form for 
further analysis after acquisition.

{\em Instrumentation particularly ideal to study nanomaterials} : 
The intensity of inelastic scattered signals from nanomaterials are 
extremely weak. The use of double/triple Monochromator causes loss of 
signals in a large amount due to presence of large number of the optical 
components in the spectrometer. To compensate this loss the use of high 
power laser is also not an ideal solution. High power density of the laser 
on the focused spot can cause unwanted phase transitions in the material.  
Use of single Monochromator with notch filter is an intelligent solution 
to above problems. With multichannel CCD detector one can use a very long 
integration time to collect very weak Raman signal from the sample for a 
wide spectral window.

{\em Drawback of the present set-up} :  Besides various advantages as 
discussed 
in this section, the above Raman set-up has the following disadvantages 
1. 
Resolution of single stage Monochromator is poor compared to double/triple 
stage Monochromator. 2. Notch filter cut off is at 150 cm$^{-1}$. Using 
this 
instrument it is not possible to record Raman spectrum below this 
wavenumber.  Thus, this set-up is not appropriate for studying acoustic 
phonons in semiconductors and optical/acoustic phonon modes in metallic 
systems. 3. The notch filter needs very careful handling. It is extremely 
sensitive to atmospheric conditions, humidity and temperature.
 
\section{Calibration}
	
The first step after installation of the instrument is to calibrate and to 
study a few standard samples.

\subsection{Calibration of the System}

\begin{figure}
\centerline{\epsfxsize=3.25in\epsffile{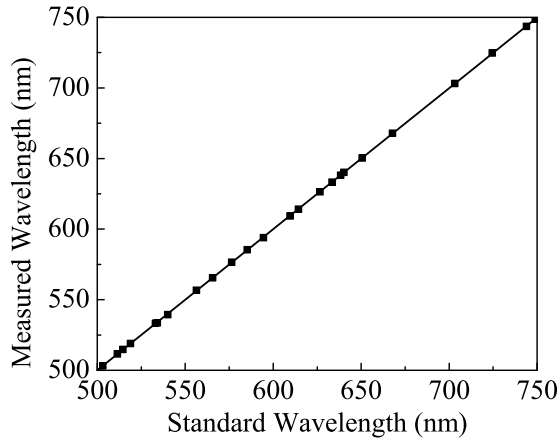}}
\caption{Plot of observed and standard Neon peaks showing linear 
correlation}
\end{figure}

The calibration of the Monochromator has been 
carried out using a Neon lamp. The entrance slit of the Monochromator was 
opened 50 $\mu$m for the calibration. At this slit width the 
resolution of the instrument is 0.03 nm (1 cm$^{-1}$).  
The spectrum was obtained in the wavelength range 
500 nm-750 nm. A plot of the measured wavelength  and the standard 
wavelength is shown in Fig. 6. The linear correlation between these two 
measures the accuracy of the measurement by the Monochromator. 

\subsection{Standard Sample Study: Solid And  Liquid}
 
Subsequent to the installation and 
calibration of the instrument a number of standard samples have been 
studied, which 
include solids and liquids. It is important to state here what we measure. 
Given a sample, its Raman spectrum is shown as a plot of intensity of 
scattered signal vs. Raman shift. Unless otherwise mentioned, in all 
figures the filled squares and the bold line represent the experimental 
data points and the fitted curve, respectively. 
For all the measurements, reported in this article, we have used 
the entrance slit width of the Monochromator as 50 $\mu$m.

\begin{figure}
\centerline{\epsfxsize=3.25in\epsffile{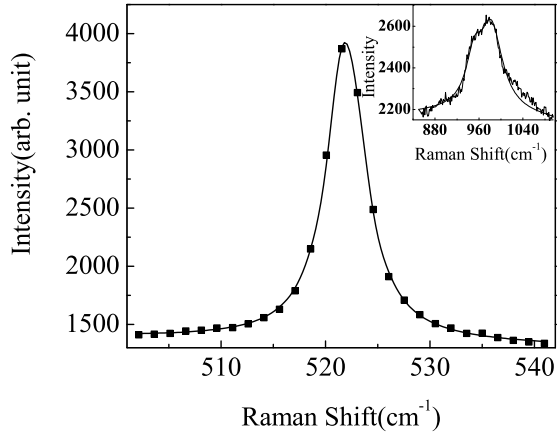}}
\caption{First order Raman spectrum of Silicon. 2nd order spectrum 
is shown in the inset.}
\end{figure}

{\em Silicon}: We have chosen silicon as a standard solid sample. Raman 
spectrum 
of a polished Silicon wafer (001) is shown in Fig. 7. The spectrum was 
taken with 80 mW power of the laser  at laser head and 50 $\mu$m 
slit-width. 
By fitting the experimental data points with Lorentzian line shape the 
first order Raman line of Silicon is obtained at {\em 521.9 $\pm 
{0.01}$ cm$^{-1}$}. The 
expected 
value for the same from the literature is 522 cm$^{-1}$  \cite{Bliz:book}. 
From 
the data analysis the width of the peak at half maxima (FWHM) is 4.4 
cm$^{-1}$. 
Weak  2nd order  TO mode of Si, measured by our instrument,  has been 
shown in inset of Fig. 7. [Zig-zag line is the measured spectrum]

\begin{figure}
\centerline{\epsfxsize=3.25in\epsffile{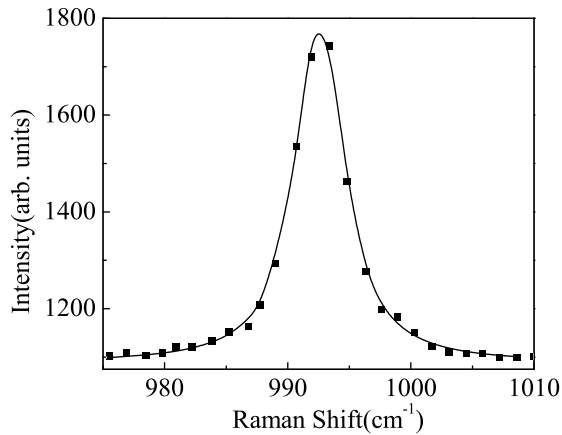}}
\caption{Raman spectrum of Benzene.}
\end{figure}

{\em Benzene} : Raman spectrum of Benzene in the range 975 cm$^{-1}$-1010 
cm$^{-1}$ is 
shown in Fig. 7. We have fitted this peak with a Lorentzian and the peak 
position is obtained at {\em 992.3 $\pm
{0.03}$ cm$^{-1}$}.  The corresponding peak due 
to 
C-C 
vibration is expected at 992.36 cm$^{-1}$ \cite{Grassmann:1933}. Thus, our 
measurement is in good agreement with the data available in the 
literature.  

Equipped with a calibrated instrument, tested with standard samples, we 
now 
move on to demonstrate how we can use it to study nanomaterials. 

\section {Studies on Nanomaterials}
In the introduction of this article we said that low dimensional systems, 
where the quantum confinement appears at the nanoscale can be of 
dimensions - two, one  or zero. Below, we provide an example of each of 
these low dimensional systems. Most of the samples have been prepared by 
us. We investigate the optical properties using PL and Raman measurements and 
point out what can be achieved by such studies. 

\subsection{Zero Dimensional System} 

{\em Ge implanted in Si:}
Semiconductor doped 
glasses show interesting optical properties as a result of quantum 
confinement of electron and hole wave functions in the embedded 
semiconductor nanocrystals \cite{Ekimov:1982,Brus:1986}. Light 
emission from Ge nanocrystals embedded in SiO$_2$ matrix is becoming an 
expanding field of interest \cite{Maeda:1991} because of their potential 
use i. as optoelectronic emission devices directly coupled with Si 
integrated circuits \cite{Shimizu-Iwayama:1993} and ii in memory device 
applications \cite{J.von:1999}. A lot of effort is being made in order to 
improve the performance of these devices. In this context, the study of 
optical properties, as we shall see, has a role to play. We have studied 
the optical properties of  Ge clusters in Si/SiO$_2$ matrix, prepared by 
low 
energy ion-implantation \cite{Yamamoto:2000}.  This technique is 
considered as one of the promising tools to synthesize embedded 
nanocrystals, as it can control the size distribution and depth profile of 
the particles within the matrix efficiently. ${}^{74}$Ge$^{+}$ ions are 
implanted in 
SiO$_2$ layer of Si wafer at room temperature with the energy 150 keV and 
with the dose 3$\times10^{16}$ ions per cm$^{2}$. Ge clusters are formed 
within the matrix 
by neutralization of charges\cite{Yamamoto:2000}.

\begin{figure}
\centerline{\epsfxsize=3.25in\epsffile{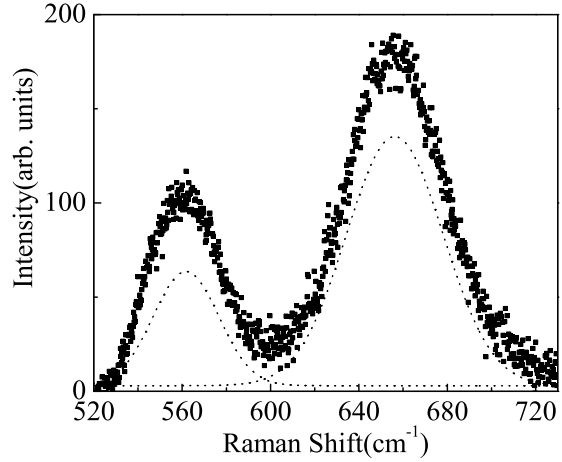}}
\caption{Photoluminescence spectrum for Ge in  Ge ion implanted 
Si/SiO$_2$ sample.}
\end{figure}

The PL spectrum of the as-implanted sample is shown in Fig. 9. The PL 
peaks appear at 562 nm and 656 nm. The first PL peak can be attributed to 
defect states originated due to implantation process \cite{Kim:1999}, 
whereas, the second peak can be due to confinement of the electron in 
oxidized Ge particles of very small size \cite{Okamoto:1996}. The 
detailed discussion on individual components are available in these 
references. Thus, such 
studies helps us to search the origin of PL in a system. For PL 
measurements we first take multiple acquisition for the whole range and 
then glue the spectra to get the PL spectrum from the sample.
 
\begin{figure}
\centerline{\epsfxsize=3.25in\epsffile{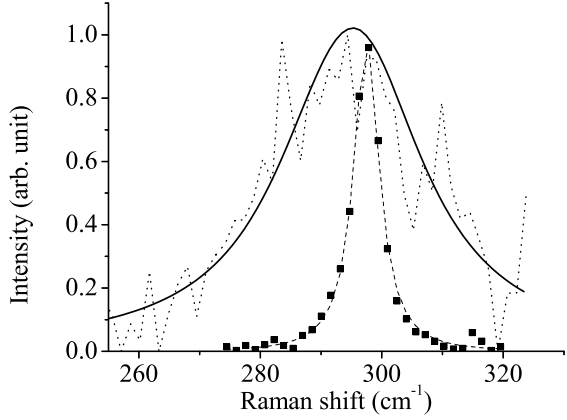}}
\caption{Raman spectrum from Ge ion implanted Si/SiO$_2$ sample. The
dotted
line represents experimental data. The bold line is the fitting of the
data with phonon confinement model. {\em The Raman spectrum of bulk Ge is
shown by filled black squre in the same graph (with a Lorentizan fitting 
by dashed line) to clarify the asymmetry and lower frequency 
shift in the Raman line for the nanoparticles.}} 
\end{figure}

In Fig. 10 the dotted line represents the 
measured Raman signal from this as-implanted sample. The integration time 
for this measurement is 300 sec. We have studied confined optic phonons in 
these Ge  nanoparticles and measured the size of the particles from the 
analysis. 
The Raman scattering of Ge particles for as-prepared 
sample shows not only broadening but also asymmetry in line shape with a 
shift towards low frequency side 
compared to that for corresponding bulk materials (Fig. 10). The 
broadening 
of the spectra is due to the amorphous nature of Ge clusters in 
as-implanted sample. The asymmetry arises due to optical phonon 
confinement. In 
bulk crystals, the phonon eigenstate is a plane wave and the wavevector 
selection rule requires q$\approx$0. In contrast, here the spatial 
correlation 
function of the phonon becomes finite due to its confinement in the 
nanocrystal and hence the q$\approx$0 selection rule is relaxed 
\cite{Campbell:1986}. The Raman spectrum $I(\omega)$ due to 
this 
confined optic phonon is given by,

\begin{equation}
I(\omega)=A \int\frac{d{\bf q}{\vert C(0,{\bf q})\vert}^{2}}{\left[\omega-
\omega({\bf q})\right]^{2} + \left(\Gamma_{0}/2\right)^{2}}
\end{equation}

\noindent
where  $\omega({\bf q})$ and $\Gamma_{0}$   are the 
phonon dispersion curve and the natural line width (FWHM) of the 
corresponding bulk materials, $C(0,{\bf q})$   is the Fourier coefficient 
of  phonon 
confinement function. {\em $A$ is an arbitrary constant.}  For 
nanoparticles it has been shown that the phonon 
confinement function, which fits the experimental data best is 
$W({\bf r},L)=\exp{(\frac{-8\pi^{2}r^{2}}{L^2})}$, 
the square of the Fourier coefficient of which is given by ${\vert 
C(0,{\bf q})\vert}^{2} \cong \exp{-\frac{q^{2} L^{2}}{16\pi^2}}$ . L is 
the 
size of the particles. The integration in Eqn. 5 must be performed over 
the 
whole Brillouin zone [45]. Since the optical phonon dispersion curves are in general not flat, phonon 
confinement
results in a peak shift (usually to lower frequency if frequency is a
decreasing function of wavevector in phonon dispersion curve) and
asymmetric broadening of the Raman line. Using 
the above analysis the fitted curve 
has been shown by the bold line in Fig. 10.  The size of the particles, 
estimated from Raman measurement, is 3.0 $\pm {0.5}$ nm.  Using 
Raman and PL 
measurement 
simultaneously, we have demonstrated both phonon and electron confinement 
in Ge nanoparticles embedded in SiO$_2$ matrix. {\em It has been 
shown in the literature that the above defect states 
are quenched after annealing the sample at 1100 $^o$C
for 30 mins. in hydrogen atmosphere
\cite{Kim:1999,Okamoto:1996,Maeda:1991}.The detail 
study on our sample with annealing, which reveals the origin of PL in 
these materials and 
role of defect states, is beyond the scope of this article and  will be 
reported 
elsewhere.}

\subsection
{One Dimensional System}

{\em The Carbon nanotube:}

\begin{figure}
\centerline{\epsfxsize=3.25in\epsffile{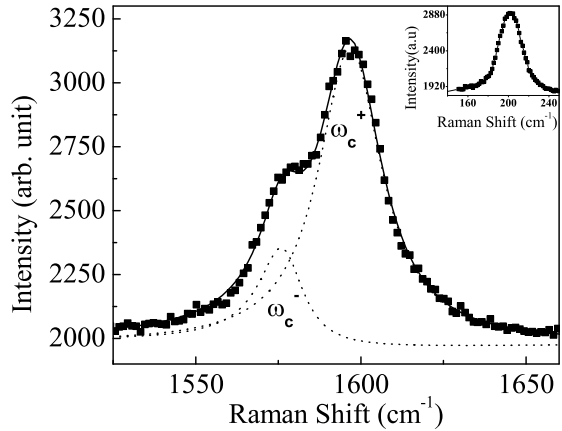}}
\caption{Dots show vibrational  G band of single walled metallic carbon 
nanotube. The solid line  shows the fit using 
$I(\omega)=I(G^+)+I(G^-)$, a 
sum of G$^+$ mode and G$^-$ mode, respectively. The radial mode is shown 
in the inset} 
\end{figure}

In 1991 Iijima  
discovered a new allotrope of carbon \cite{Iijima:1991}, carbon nanotube, 
in which the graphite sheets rolled up to form hollow cylinders of few 
nanometer in diameter but upto a thousand times as long. These nanotubes 
have revolutionized low dimensional Physics and are utilized extensively 
in state of art nanoscale research, as mentioned in Section I. The 
earliest observations were of micron-long multi walled carbon nanotubes 
found in the soot by carbon arc method\cite{Kratschmer:1990}. We have 
prepared single walled metallic carbon nanotube (SWNT)by thermal 
decomposition 
of Acytelene gas in nitrogen atmosphere at 700 $^o$C on 
porous silicon 
containing iron particles.  Here we demonstrate the capability of our 
system to characterize milligram quantities of carbon nanotubes. Fig. 11 
shows the Raman spectrum of carbon nanotube recorded in our instrument. In 
this figure, the multi-peak tangential breathing mode (G-band) and low 
frequency radial mode (in the inset) are the characteristics of metallic 
single wall carbon nanotube. The Raman-allowed tangential G-mode in 
graphite is observed at 1582 cm$^{-1}$.  Unlike graphite, the tangential G 
mode 
in a nanotube gives rise to multipeak features, known as G-band 
\cite{Brown:2001}. An analysis can been carried out considering the two 
intense features in G-band, which originate from the symmetry breaking of 
the tangential  vibration when the graphite sheet is rolled to make a 
cylindrical shaped nanotube.  The most intense G peak, G$^{+}$ band, is 
for the 
atomic displacements along the tube axis and G$^{-}$ band is caused by the 
curvature of the nanotube, which softens the tangential vibration in the 
circumferential direction. In Fig. 11 the peak positions for G$^{+}$ band 
and G$^{-}$ 
band appear at 1596.8 $\pm {0.20}$ cm$^{-1}$  ($\omega_{G}^{+}$)  
and 576.8 $\pm {0.20}$ cm$^{-1}$ ($\omega_{G}^{-}$), respectively. 
For 
metallic SWNT the $G^{-}$ band is broadened due to presence of free 
electrons 
in nanotubes with metallic character. We have fitted this G$^{-}$ feature 
by Briet-Wigner-Fano (BWF) line shape, which is given by

\begin{equation}
I(\omega)=I_{0} 
\frac{[1+(\omega-\omega_{BWF})/q\Gamma]^{2}}{1+[(\omega-\omega_{BWF})/\Gamma]^{2}}
\end{equation}

\noindent           
where -1/q is a measure of the interaction of the phonon with a 
continuum of states and $\omega_{BWF}$ is the BWF peak frequency at 
maximum 
intensity $I_{0}$ \cite{Brown:2001}. From the fit the value of -1/q has 
been 
found to be 0.14, which is close to the value reported in the 
literature. The higher wavenumber G$^{+}$ band has been fitted with 
Lorentzian line shape. The splitting $\Delta\omega_{G} 
=\omega_{G}^{+}-\omega_{G}^{-}$ can be used for diameter characterization 
\cite{Jorio:2003}. In Fig. 11 these two fitted curves are shown by 
the dotted 
lines and the net fitted curve is shown by the solid line.  

\vspace{0.5cm}

{\em Porous silicon :}

\begin{figure}
\centerline{\epsfxsize=3.25in\epsffile{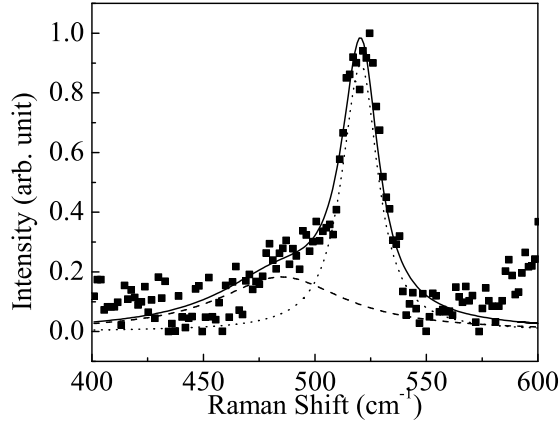}}
\caption{Dots show the Raman spectrum of as-prepared porous silicon. The 
solid line  shows the fit using $I(\omega)=I_{d}(\omega)+I_{c}(\omega)$, 
a sum of phonon 
confinement effect (dotted line) and disordered spectrum (dashed line), 
respectively.} 
\end{figure}

The recent 
observations of highly efficient visible photo- and electro luminescence 
at room temperature from electrochemically etched porous silicon (PS) have 
stimulated a lot of excitement mainly due to the possibility of its use in 
Si-based device application and in designing different  sensors 
\cite{Hirschman:1996}. PS has been prepared by electrochemical etching of  
p-type Si wafer of $<$100$>$ orientation in 48$\%$ HF:ethanol (1:1) 
solution, keeping current density between 10 and 15 mA/cm$^{2}$.
PS can be approximated as quantum wires. Formation 
of PS on p-type silicon is a self adjusting process due 
to a geometrical quantum wire effect associated with thin silicon walls 
remaining between pores after electrochemical etching. A sponge like 
morphology of PS has been characterized in terms of 
fractals (an intricate infrastructure consisting of a hierarchy of 
pores within pores) within certain finite length scale \cite{Chuang:1989}. 
Both, quantum confinement of electrons in the nanowire and the contribution of 
defect/disordered states present in this system are important to  explain 
the origin of luminescence in PS \cite{Roy:1994}.The 
presence of disordered/amorphous phase has been reported earlier from 
EPR measurements \cite{Prokes:1992}. Fig. 12 shows the Raman spectrum 
of PS. Keeping in mind the 
possibility of presence of disordered/amorphous Silicon (Si) together with 
the crystalline nanowire of Si, we have fitted the observed line shape  
with a combined line shape $I(\omega)=I_{c}(\omega)+I_{d}(\omega)$, where 
$I_{c}(\omega)$ is the confinement Raman line shape (dotted line), given by 
Eqn. 5 and 
$I_{d}(\omega)=B\Gamma_{d}[\omega-\omega_{d})^{2}+\Gamma_{d}^{2}]^{-1}$ 
is the Lorentzian line shape (dashed line), which arises due 
to the disordered/amorphous component. Here $\omega_d$ and $\Gamma_{d}$ 
are the phonon frequency and half width at half maximum (HWHM), 
respectively for the disordered component and B is a constant. The best fit 
line is shown by the bold line in Fig. 12. In the above exercise of 
nonlinear least square fit of the data with I($\omega$), the fitting 
parameters are $L$, $A$ (both from Eqn. 5), {\em $\omega _{d}$, 
$\Gamma_{d}$} and $B$. 
The phonon dispersion curve for bulk silicon, $\omega$ ({\bf q}), is 
taken 
from Ref.\cite{Roy:1994}. The diameter of the quantum 
wire $L$, estimated from the above Raman analysis, is 40 $\pm
{5}$ \AA.
Simultaneous 
PL and Raman measurements on porous silicon can be used to study the origin 
of PL in this material \cite{Roy:1994}. 

\subsection
{Two Dimensional System}

{\em 2D InGaAs quantum layer:}

\begin{figure}
\centerline{\epsfxsize=3.25in\epsffile{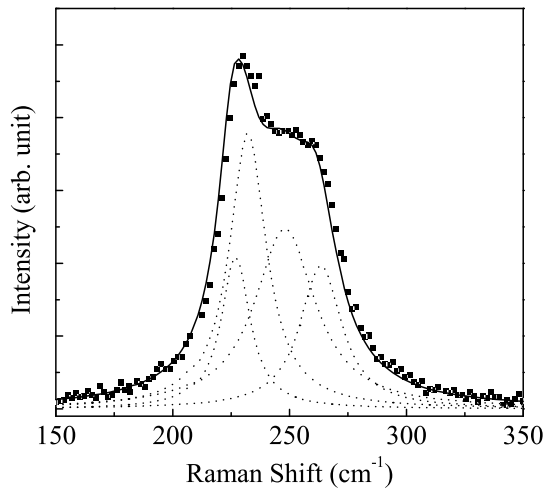}}
\caption{Raman spectrum of InGaAs 2D layer grown on InP substrate. The 
dotted lines are four components of LO and TO modes of GaAs and InP. The 
solid line is the best fit to the data using the above components.} 
\end{figure}

With the rapid 
development in modern micro/nano electromechanical and VLSI devices it is 
important to understand the structural and elastic properties of these 
semiconductor layers\cite{Oishi:1997,Ochiai:1995}. When a 
superlattice is formed by growing thin layers of materials on a substrate, 
the differences in their thermal expansion coefficients and lattice 
constants in these two materials lead to stresses and strains within the 
layers\cite{Pollak:book}. Measurement of Raman spectra in these 
superlattices provides the useful information about the microstrain in 
semiconductors under stresses \cite{Cerderia:1972}. Here we have reported 
Raman spectrum of as-grown 1$\mu$m thick In$_{0.53}$Ga$_{0.47}$As layer 
on InP 
substrate. Thin film In$_{0.53}$Ga$_{0.47}$As has been grown on low 
pressure MOCVD 
(Metal-Organic Chemical Vapour Deposition) system (CVD Equipment 
Corporation) at 629 ${}^{o}$C temperature and under a growth pressure of 
100 
Torr. Flow rate of carrier (H$_{2}$) gas was 1.3 lit/min. In this case 
n$^{+}$ - InP 
was used as the substrate and 0.5 $\mu$m undoped InP buffer layer was 
grown on 
it. 

In Fig. 13 we have shown Raman spectrum of as-grown 
In$_{0.53}$Ga$_{0.47}$As film on 
InP substrate. Expecting the presence of both LO and TO modes for 
InAs and GaAs Raman spectrum between 150 to 270 cm$^{-1}$ is 
decomposed into 4  Lorentzians.  For the nonlinear least square fit of 
the data, we have used the FWHM  and peak intensities as free fitting 
parameters. The peak positions are kept within $\pm$ 5 cm$^{-1}$ of the 
bulk LO and TO phonon modes for GaAs and InAs.
Peak A appears between 226.7 cm$^{-1}$  and 229.4 cm$^{-1}$ (InAs-like TO 
mode), 
peak B - at 241.5 cm$^{-1}$ (InAs-like LO mode), peak C - at 254.3 
cm$^{-1}$ 
(GaAs-like TO mode) and peak D between at 266.0 cm$^{-1}$ 
(GaAs-like LO mode). In 
the above fitting procedure it is difficult to comment on Raman shift of 
peak B and C distinctly, as they are very close to each other. We have 
estimated the stress within the layers from the shift in Raman peak 
positions of GaAs-like longitudinal optical (LO) phonons in the systems. 
Though at room temperature In$_{0.53}$Ga$_{0.47}$As/InP is almost 
perfectly lattice 
matched, at the growth temperature (629${}^{o}$C), 
In$_{0.53}$Ga$_{0.47}$As has a lattice 
constant less than that of InP and the layer is under compression 
\cite{Yagi:1983}. GaAs-like Raman mode for lattice matched 
In$_{0.53}$Ga$_{0.47}$As/InP appears at 270 cm$^{-1}$ 
\cite{Cerderia:1972}. 
The phonon 
frequencies of the same mode in our samples, as obtained by fitting the 
Raman spectrum is 266.0 $\pm {0.56}$ cm$^{-1}$. The shift in the Raman 
line 
from its 
value in 
lattice matched sample can be attributed to the biaxial stress in the film 
due to lattice mismatch. Following \cite{Cerderia:1972} \& 
\cite{Yagi:1983} 
the frequency of the optical phonons in the presence of biaxial stress can 
be expressed by

\begin{equation}
\Omega_{LO}=\Omega_{LO}^{0}-\eta_{H}\sigma-\frac{2}{3}\eta_{S}\sigma
\end{equation}   
         
where $\Omega_{LO}^{0}$ is the phonon frequency without stress and

\begin{equation}
\eta_{H}=-(p+2q)(S_{11}+S_{12})\Omega_{0}/6{\omega}_{0}^{2}
\end{equation}

and

\begin{equation}
\eta_{S}=(p-q)(S_{11}-S_{12})\Omega_{0}/2{\omega}_{0}^{2}
\end{equation}         

\noindent
Here $S_{11}$ and $S_{12}$ are the elastic stiffness constants, $p$ and 
$q$ are the 
optical deformation constants. In above Eqns 8 \&
9, optical phonon frequency, $\Omega$, is in unit of cm$^{-1}$ and 
$\omega_{0}$ is  in sec$^{-1}$. The values of $p+2q$, $p-q$, 
$S_{11}+2S_{12}$ and 
$S_{11}-S_{12}$, 
$\omega_{0}$, are 
tabulated in Table 2 (from \cite{Cerderia:1972}). In above Eqns 8 \& 
9,  $\Omega$'s are in unit of cm$^{-1}$ and $\omega_{0}$ is  in sec$^{-1}$ 
The compressive stress in 
the film is then given by the expression 
\begin{equation}
\sigma=-0.4{\partial}{\Omega}_{LO} GPa
\end{equation}

\noindent
where ${\partial}{\Omega}_{LO}={\Omega}_{LO}^{0}-{\Omega}_{LO}$. $\sigma$ 
is 
positive for tension and negative for compression.The stresses within the 
layers in the samples, estimated from the Raman measurements, is 
1.6$\pm$0.2 GPa.

\begin{widetext}
\begin{table}
\renewcommand{\tabcolsep}{.9pc}
\begin{tabular}{|c|c|c|c|c|c|}\hline
$\Omega_0$ (s$^{-1}$) & $S_{11}+2 S_{12}$ (Pa$^{-1}$) & $S_{11}-S_{12}$ 
(Pa$^{-1}$) & (p+2q)/6$\omega_{0}^{2}$ & (p-q)/2$\omega_{0}^{2}$ \\ 
\hline \hline
0.511 $\times 10^{14}$ & 0.445 $\times 10^{-11}$ & 1.54 $ \times 10^{11}$ & 0.2 &
-0.1 \\ \hline
\end{tabular}
\caption{Elastic stiffness constant, optical deformation constant for
InGaAs}
\end{table}
\end{widetext}

\section {Conclusion}
The ability to control, manipulate and design materials in the nanometer 
scale is one of the major technology drivers of the $21^{st}$ century. 
Thus, to 
use these materials one needs to understand their basic physical 
properties in details. We have described design and fabrication of a 
low-cost spectrometer with home-built collection optics for advanced 
research projects, specially ideal to study optical and vibrational 
properties of nanomaterials. The low cost open electrode CCD detector 
enables us to acquire weak signals with large integration time. High 
throughput of a single stage Monochromator also helps us to detect very 
weak signals from nanomaterials. The system can be easily modified to a 
MicroRaman set-up, with a microscope attachment. 

To establish the performance of our instrument for studying present state 
of art materials we have worked on 0D system (Ge nanoparticles), 1D system 
(Si nanowire, Carbon nanotube) and 2D system (InP-InGaAs thin film). The 
spectrometer could successfully record very weak Raman signal from these 
samples. Different techniques to analyze the data to get useful 
information about the sample have been briefly described in this article. 
Using same set-up we could also measure PL from the samples. Simultaneous 
PL and Raman measurements are important to study these new materials. PL 
experiments help in understanding the nature of luminescent states, 
whereas Raman scattering probes the underlying quantum network. 	

A wide range of studies can be performed using the same instrument.  For 
example, Surface Enhanced Raman Scattering (SERS) \cite{Kneipp:1997} is a 
useful technique 
resulting in strongly increased Raman signals from molecules, which have 
been attached to nanometer scale metallic structure. It is an unique 
technique to probe individual molecules.  Our spectrometer is capable of
carrying out this single molecule spectroscopy. To begin with, we have 
probed 
molecular vibrations in single molecule of methylene blue dye absorbed on 
silver sol by SERS.  	

To conclude, we have described a set-up for Raman and PL measurements, 
which can be ideally used in research as well as in any graduate level 
course on nanomaterials, where the importance of spectroscopic 
investigation is adequately emphasized. Through the differences that arise 
in the spectroscopic effects, we can mark and measure the unique features 
inherent to the nanostructures within the sample. The examples provided in 
this article illustrate quite clearly how our Raman PL set-up can 
efficiently probe the nano -aspects, so important in modern-day 
technology. We hope that this article will motivate researchers to make 
use of our set-up for appropriate purposes in future.

\section{Acknowledgements}
AR thanks Department of Science and Technology (DST) in India for 
financial 
support and Dr. S. Kar for fruitful suggestions and discussions. Authors 
thank Nuclear Science Centre, Delhi, India, for help in preparing the 
Ge implanted SiO$_2$ sample, Prof. D.N. Bose for MOCVD grown InGaAs 
quantum well sample and M.S. Morrison for technical help with the 
set-up.


\end{document}